\documentclass[11pt]{article}

\usepackage{amsthm}
\usepackage{graphicx}
\usepackage{epstopdf}
\usepackage{amsmath,amstext,amssymb,amsfonts,verbatim}
\usepackage{xcolor}
\usepackage{hyperref}
\usepackage{algorithmic,algorithm,rotating}

\hypersetup{colorlinks=false,linkbordercolor=red,linkcolor=blue,pdfborderstyle={/S/U/W 1}}

\begin{document}

\title{\vspace{-2cm}Quantifying Loss of Information in Network-based Dimensionality Reduction Techniques 
}


\author{Hector Zenil, Narsis A. Kiani\thanks{H.Z. and N.K. contributed equally to this work.}, Jesper Tegn\'er\\
Unit of Computational Medicine, Centre for Molecular Medicine and\\
Science for Life Laboratory (SciLifeLab), Department of Medicine\\
Karolinska Institutet, Stockholm, Sweden \\
              {\{hector.zenil, narsis.kiani, jesper.tegner\}@ki.se}          }

\date{}

\maketitle

\begin{abstract}

To cope with the complexity of large networks, a number of dimensionality reduction techniques for graphs have been developed. However, the extent to which information is lost or preserved when these techniques are employed has not yet been clear. Here we develop a framework, based on algorithmic information theory, to quantify the extent to which information is preserved when network motif analysis, graph spectra and spectral sparsification methods are applied to over twenty different biological and artificial networks. We find that the spectral sparsification is highly sensitive to high number of edge deletion, leading to significant inconsistencies, and that graph spectral methods are the most irregular, capturing algebraic information in a condensed fashion but largely losing most of the information content of the original networks. However, the approach shows that network motif analysis excels at preserving the relative algorithmic information content of a network, hence validating and generalizing the remarkable fact that despite their inherent combinatorial possibilities, local regularities preserve information to such an extent that essential properties are fully recoverable across different networks to determine their family group to which they belong to (eg genetic vs social network). Our algorithmic information methodology thus provides a rigorous framework enabling a fundamental assessment and comparison between different data dimensionality reduction methods thereby facilitating the identification and evaluation of the capabilities of old and new methods.\\


\end{abstract}

\section{Introduction}

The advent of high-throughput genomics technologies has made available large quantities of data, transforming molecular biology into a remarkably data-rich science. Each passing year sees an increase in the use of high-dimensional data to probe everything from gene regulation and the evolution of genomes to the individual genetic profile of complex disease development. Life scientists now find themselves having to cope with huge data sets, and face challenges extracting and interpreting the wealth of information hidden in these data. 

Representing data in a well-studied formal structure is ideally suited to follow-up analysis and to addressing many of the questions arising from the interpretation of large scale data. 
Recently developed experimental and computational techniques yield networks of increased size and sophistication. The study of such complex cellular networks is emerging as a new challenge in biology. Network science is now central to molecular biology, serving as a framework for reconstructing and analyzing relations among biological units~\cite{barabasi,mendes,newman,alon}.

The characteristic combination in biology of minute observation and a large number of variables results in very dense networks, the upshot of which, from a data analysis perspective, is the so-called ``curse of dimensionality" problem~\cite{1}. Biological networks carry information, transfer information from one region to another and implement functions represented by the network's interactions. 

The visualization and analysis of such networks can pose significant challenges, which are often met by identifying the backbone of complex networks. Over the last decade, determining the vital features of these huge networks has been an intriguing topic, and continues to be a challenge. Dimension reduction methods offer a potentially useful strategy for tackling such problems. They aim to reduce the predictor dimension prior to any modeling efforts. The main aim of all these efforts is to extract a processing core from large noisy networks. Surprisingly, the amount of information lost or conserved in so doing has remained unknown or unquantified. Furthermore, there is no general framework for evaluating and comparing these methods. 

Here we propose a novel approach for studying the complexity of biological networks and for evaluating network dimensionality reduction processes, applying information-theoretic measures to detect global and local patterns. In particular, we study the rate at which information can be lost, recovered or reconstructed in reduced complex artificial and real networks, while retaining the typical features of biological, social, and engineering networks, such as scale-free edge distribution and the small-world property. We will use a more powerful measure of information and randomness than Shannon's information entropy, namely, the so-called Kolmogorov complexity $K$. $K$ has been proven to be a universal measure theoretically guaranteed to asymptotically find any computable regularity~\cite{solomonoff} in a dataset. $K$ can be effectively approximated by using lossless compression algorithms, for example. That is, compression algorithms for which decompression fully recovers the original object, with no loss of information. A good introduction to the subject may be found in~\cite{li} and~\cite{calude}. To approximate Kolmogorov complexity, we use a technique called the \textit{Block decomposition method}~\cite{zenilgraph} (or simply BDM) based on algorithmic probability~\cite{d5} and two generally employed lossless compression algorithms, Bzip2 and Deflate.

Bzip2 is an open source data compressor that uses a stack of different algorithms superimposed one atop the other starting with run-length encoding, Burrows-Wheeler or the Huffman coding, among others. We sometimes compare, strengthen or complement findings by also providing estimations of Shannon's information entropy.

While more dimension reduction techniques can be conceived of than can be thoroughly analyzed in a single paper, we provide the tools and methods with which to do so, regardless of the technique. Here, however, we compare three distinct graph dimension reduction techniques (graph spectrum, sparse graph and motif profile) and evaluate their ability to preserve the information content of the original network. These methods have been applied to different biological networks in order to understand complex cellular behaviours~\cite{3,4,6}.

The logic behind the use of motif profiles is the basic assumption that the over-representation of a certain motif in a network indicates that it has some functional importance. Thus, exploring the most frequently occurring motifs in a network may afford novel insights into the functionality of these motifs within the network. FANMOD~\cite{fanmod} has been used to find network motif profiles. The sparse networks have been obtained by applying the effective resistances sparsification method. Effective resistances sparsification has been reported to be one of the quickest sparsification methods, which keeps the backbone of a network intact~\cite{spielman}.

We compare what the three methods (see Appendix), graph spectra, graph motifs and graph sparsification-- which are clearly forms of lossy compression as the networks cannot be fully recovered--capture, and we test whether they characterize families of networks. In other words, we measure the ability of these methods to preserve key information. We show that they not only capture different properties but also preserve different amounts of information from the original objects. There were four main sources of networks to which dimensionality reduction methods and information-theoretic measures were applied  One source was tailored graphs produced specifically for this paper, such as spider graphs and co-spectral graphs. Real-world networks come from the landmark paper where network motifs for systems biology was introduced~\cite{milo}. 

Finally, from the widely-known Artificial Gene Network Series Century database (Mendes DB)~\cite{mendesdb}, a sample comsisting of two small-world networks (SW), two scale-free networks (SF) and two Erd\"{o}s-R\'enyi networks (ER) were used, all of them with 100 nodes and 200 edges. These are public data sources of well-known networks, used instead of custom-made networks in the interest of impartiality. Methods and measures were thus applied to networks that are widely available and not to networks contrived to suit the particular methods or measures applied in this paper. From now on all graphs analyzed, whether natural or synthetic, are directed, but no information regarding activation or inhibition is taken into account (since for several of them there is none).

\section{Results}

\begin{table}[!ht]
\tiny
\caption{\bf{Complexity approximation by BDM, Deflate and Bzip2 of all original graphs.}}

\begin{tabular}{|c|c|c|c|c|c|c|c|c|c|}

  \hline

 \textbf{Network} & BDM & Deflate & Bzip2 & BDM & Deflate & Bzip2 & BDM & Deflate & Bzip2 \\

  &  &  &  & / $|V|$ &  / $|V|$ &  / $|V|$ &  / $|E|$ &  / $|E|$ &  / $|E|$ \\

 \hline

 \textbf{yeast} & 1903.94 & 4014 & 1441 & 2.76 & 5.83 & 2.1 & 1.76 & 3.72 & 1.33 \\

  \hline

 \textbf{ecoli} & 1387.65 & 2250 & 859 & 3.31 & 5.38 & 2.05 & 2.67 & 4.33 & 1.65 \\

  \hline

 \textbf{leader2Inter} & 977.6 & 494 & 163 & 30.55 & 15.43 & 5.09 & 10.18 & 5.14 & 1.69 \\

 \textbf{(social net 2)} &  &  &  &  &  &  &  &  &  \\

 \hline

 \textbf{scale-free 2} & 1353.05 & 922 & 382 & 13.53 & 9.22 & 3.82 & 6.76 & 4.61 & 1.91 \\

 \hline

 \textbf{1AORInter} & 1663.13 & 938 & 416 & 17.14 & 9.6 & 4.28 & 7.84 & 4.42 & 1.96 \\

 \textbf{(protein 1)} &  &  &  &  &  &  &  &  &  \\

 \hline

 \textbf{prisonInter} & 1432.05 & 850 & 360 & 21.37 & 12.68 & 5.37 & 7.86 & 4.67 & 1.97 \\

 \textbf{(social net 1)} &  &  &  &  &  &  &  &  &  \\

 \hline

 \textbf{1eawInter} & 1136.41 & 610 & 245 & 21.44 & 11.5 & 4.62 & 9.24 & 4.96 & 1.99 \\

 \textbf{(protein 2)} &  &  &  &  &  &  &  &  &  \\

 \hline

 \textbf{1a4jInter} & 1443.11 & 966 & 428 & 15.19 & 10.16 & 4.5 & 6.77 & 4.53 & 2 \\

 \textbf{(protein 3)} &  &  &  &  &  &  &  &  &  \\

 \hline

 \textbf{scale-free 1} & 1321.49 & 958 & 410 & 13.21 & 9.58 & 4.1 & 6.6 & 4.79 & 2.05 \\

 \hline

 \textbf{small world 1} & 919.69 & 898 & 472 & 9.19 & 8.98 & 4.72 & 4.62 & 4.51 & 2.37 \\

 \hline

 \textbf{small world 2} & 919.69 & 898 & 472 & 9.19 & 8.98 & 4.72 & 4.62 & 4.51 & 2.37 \\

 \hline

 \textbf{erd\"os-r\'enyi 1} & 1062.02 & 1046 & 492 & 10.62 & 10.46 & 4.92 & 5.31 & 5.23 & 2.46 \\

 \hline

 \textbf{erd\"os-r\'enyi 2} & 1107.83 & 1046 & 500 & 11.07 & 10.46 & 5. & 5.53 & 5.23 & 2.5 \\

 \hline

 \textbf{s208} & 1172.59 & 1074 & 526 & 9.61 & 8.8 & 4.3 & 6.2 & 5.68 & 2.78 \\

 \textbf{(electric 1)} &  &  &  &  &  &  &  &  &  \\

 \hline

 \textbf{s420} & 1835.59 & 2174 & 1131 & 7.28 & 8.62 & 4.48 & 4.6 & 5.44 & 2.83 \\

 \textbf{(electric 2)} &  &  &  &  &  &  &  &  &  \\

 \hline

 \textbf{s838} & 2032.63 & 4334 & 2446 & 3.96 & 8.46 & 4.77 & 2.48 & 5.29 & 2.98 \\

 \textbf{(electric 3)} &  &  &  &  &  &  &  &  &  \\

  \hline

\end{tabular}

\begin{flushleft} Complexity of all graphs approximated by the three methods: BDM, Deflate and Bzip2, normalized by number of nodes and number of edges. List sorted by last column Bzip2 normalized by number of edges. A figure plotting these values for comparison and normalized between 0 and 1 is provided in Fig.~\ref{figure3}.
\end{flushleft}
\label{tab:table1}
\end{table}

\begin{figure}[!ht]

\begin{center}

\includegraphics[width=12cm]{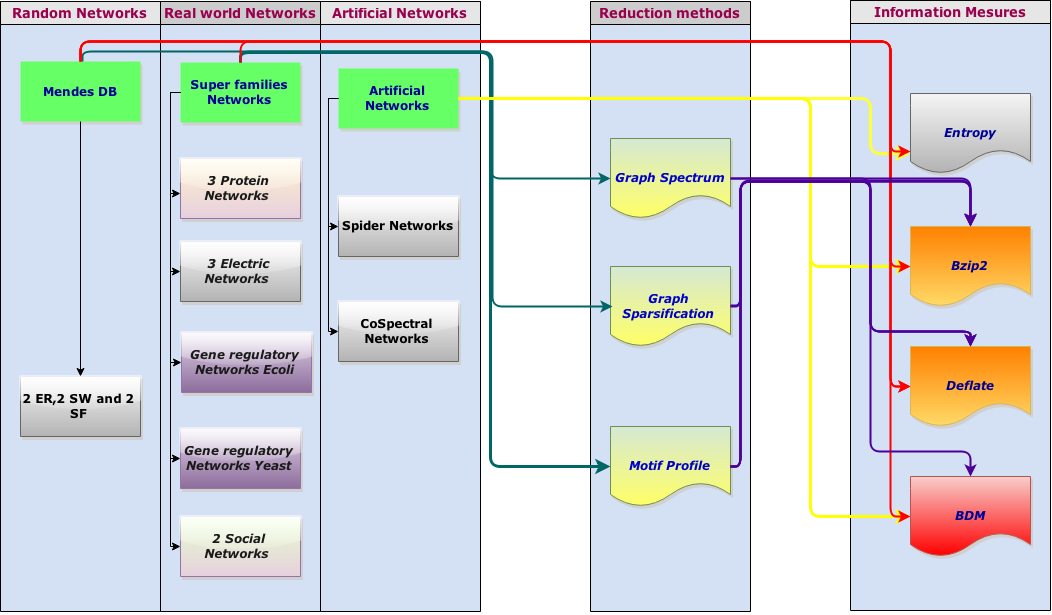}

\end{center}

  \caption{{\bf Flowchart Quantifying Loss of Information in Network-based Dimensionality Reduction Techniques}. Main results are shown in Figs.~\ref{figure3}, \ref{figure4} and \ref{figure5}.}

  \label{figure1}

\end{figure}

\begin{figure}[!ht]

\begin{center}

\includegraphics[width=11cm]{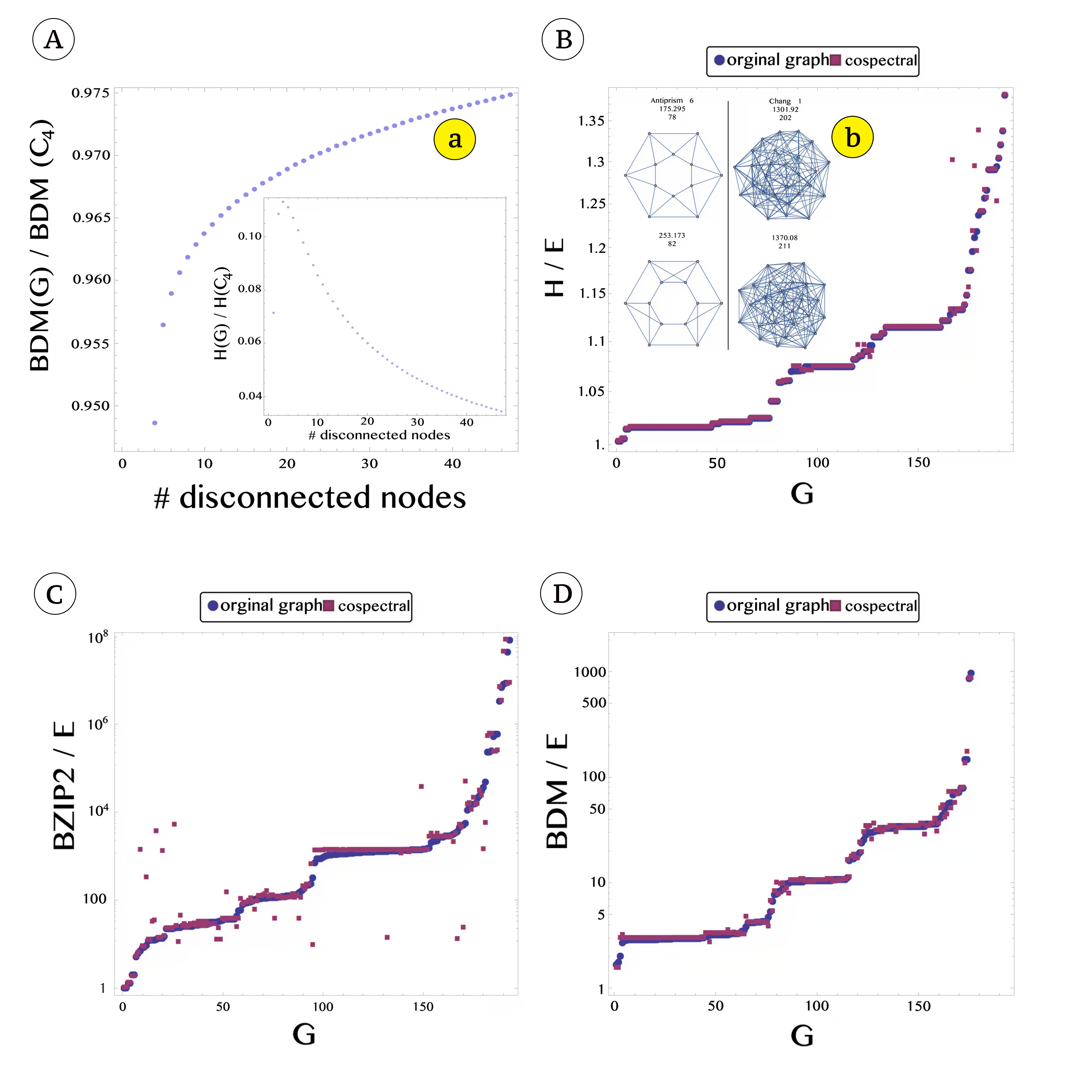}

\end{center}

  \caption{{\bf Information content of graphs and networks.}  (A) Asymptotic behavior of BDM and of (a) Shannon's entropy when adding disconnected nodes to a cycle graph of size 4 as a test of error estimation of graph entropy and graph complexity. (Bb) Examples of cospectral graphs. Entropy (B) and algorithmic complexity estimations by Bzip2 (C) and BDM (D) for a set of 180 graphs and their cospectrals.}

  \label{figure2}

\end{figure}

\begin{figure}[!ht]

\begin{center}

\includegraphics[width=13.7cm]{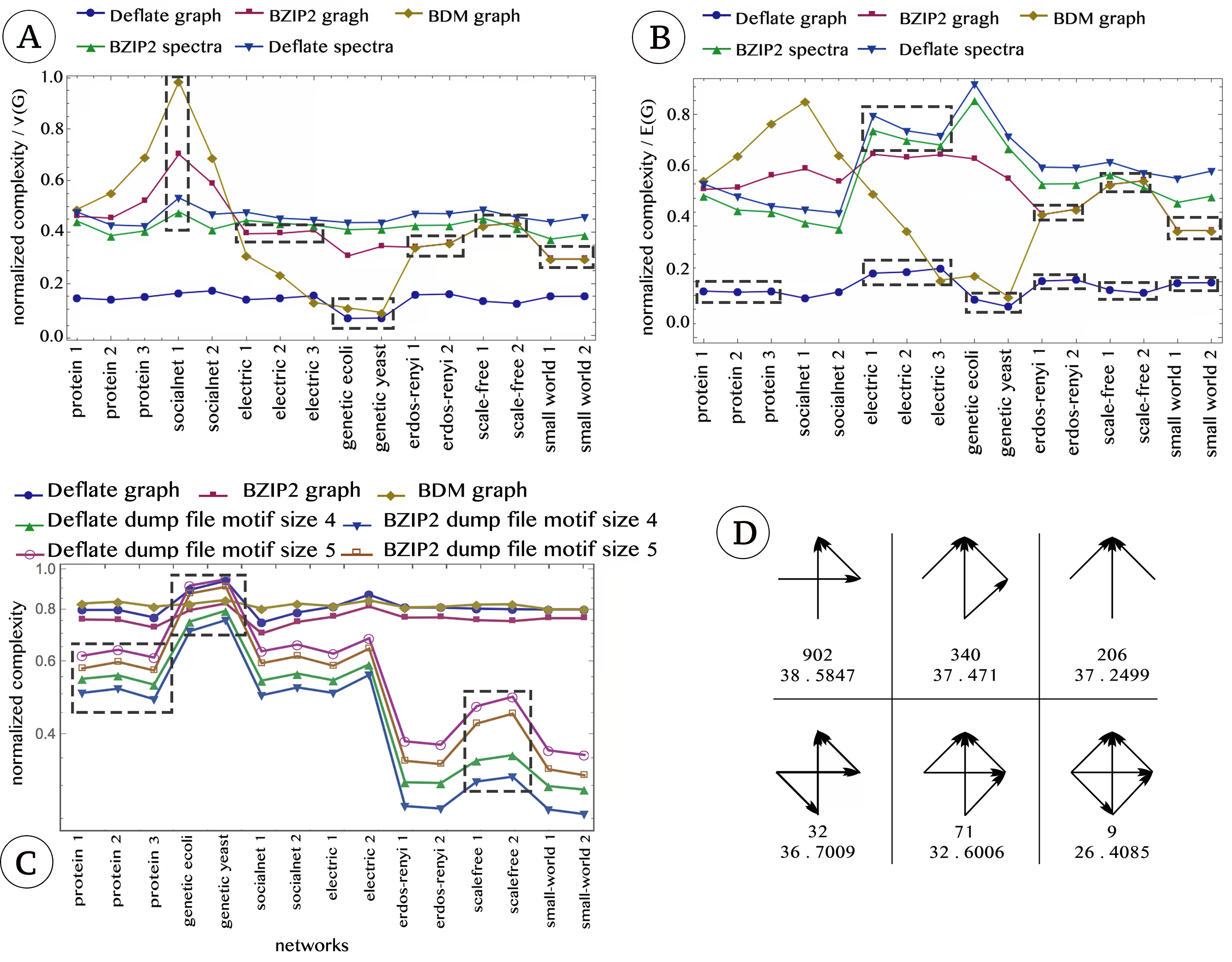}

 \end{center}

  \caption{{\bf Information content of graph spectra and graph motif analysis.} (A-B) Plots comparing the information content of 16 graphs of different types and the information content of their graph spectra approximated by Bzip2 and Deflate lossless compression, normalized by node (A) and edge count (B) in order to delete the effect of each and ascertain which may be driving the measures. (C) Plot of the same set of networks, comparing the original information content of the graphs and the information content of the number and type of graph motifs normalized between 0 and 1. (D) Shows the complexity by Deflate (top value) and Entropy (bottom value) of each of the 6 non-isomorphic graph motifs of size 4. Data points were joined only for ease of reading. Gray rectangles with dashes mark patterns that different methods  pick up from the network signatures after and before size reduction. 
  }

\label{figure3}

\end{figure}

\begin{figure}[!ht]

\begin{center}

\includegraphics[width=13cm]{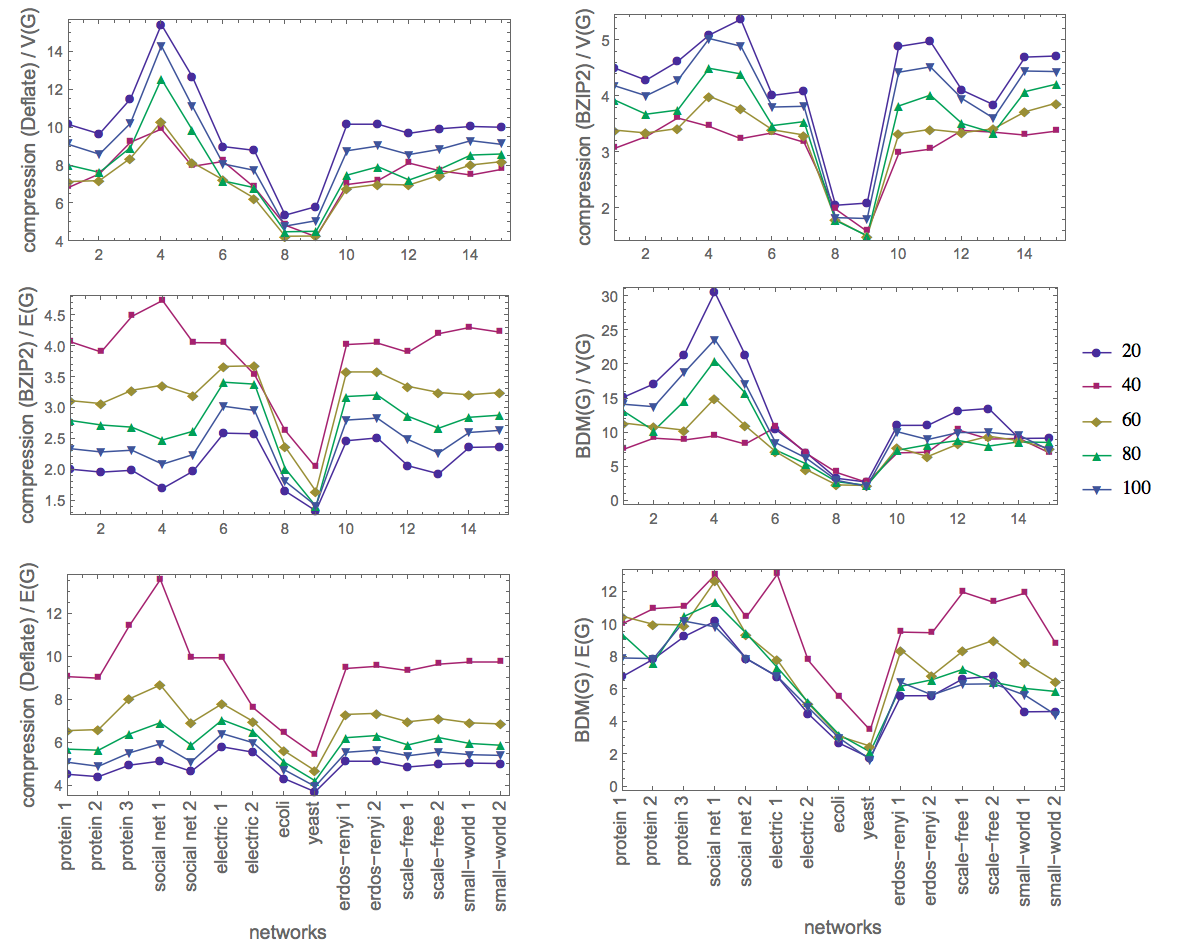}

 \end{center}

  \caption{{\bf Information content progression of \textit{spectral sparsification}.} Information loss after keeping from 20 to 80\% of the graph edges (100\% corresponds to the information content of the original graph). For comparison purposes we considered normalized complexity values by both number of edges and nodes for Deflate, Bzip2 and BDM. Data points were joined only for ease of reading.}

  \label{figure4}

\end{figure}

\begin{figure}[!ht]

\begin{center}

\includegraphics[width=10cm]{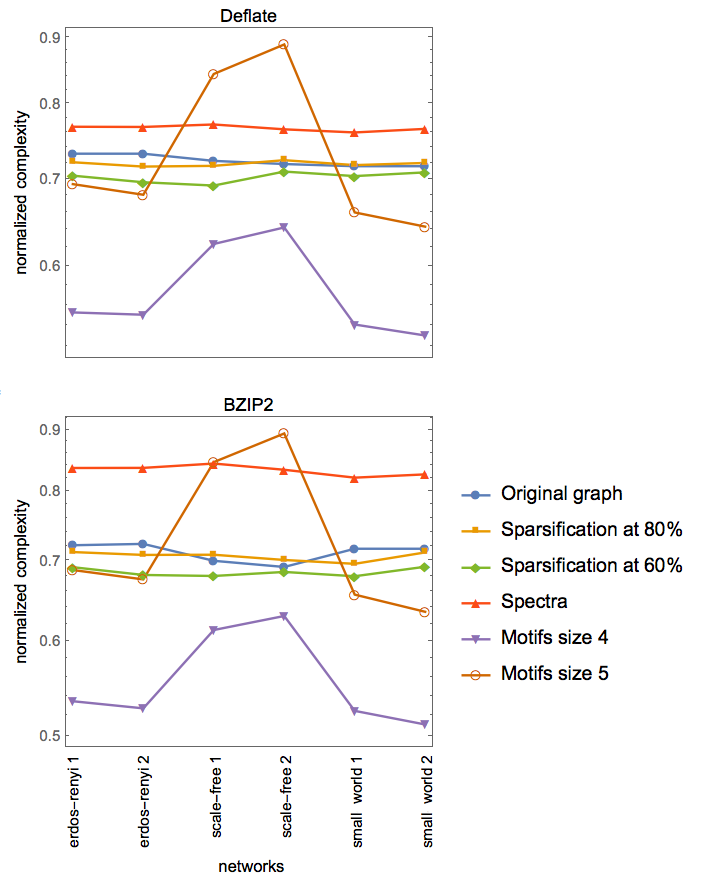}

 \end{center}

  \caption{{\bf Plot comparing all methods as applied to 4 artificial networks.} The information content measured as normalized complexity with two different lossless compression algorithms was used to assess the sparsification, graph spectra and graph motif methods. The 6 networks from the Mendes DB are of the same size and each method displays different phenomena. Data points were joined only for ease of reading.}

  \label{figure5}

\end{figure}

\begin{figure}[!ht]

\begin{center}

\includegraphics[width=7cm]{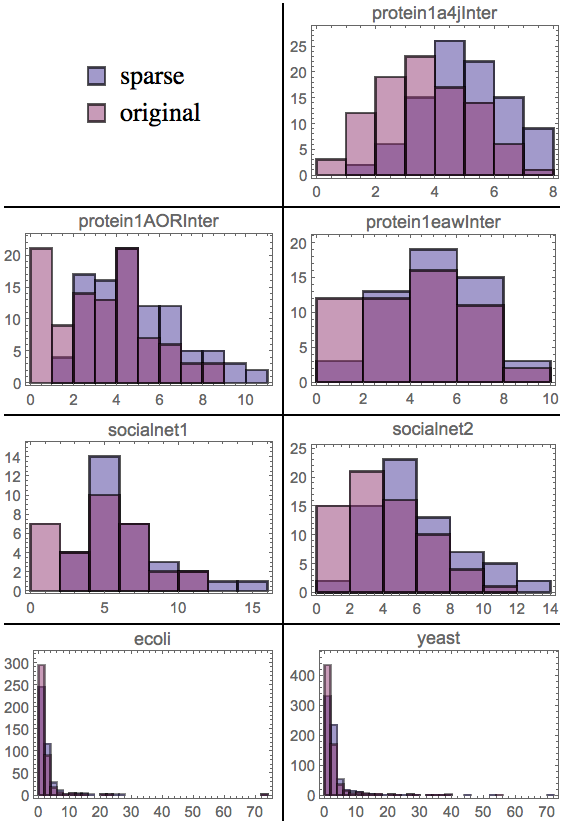}

\end{center}

  \caption{Except in the case of one of the protein networks, the graph spectrum sparsification method preserves the degree distribution of the original graph relatively well, keeping 80\% of the original edges in all real-world networks considered in this paper.}

  \label{figure6}

\end{figure}

\subsection{Information content of networks}

The complexity of biological networks may be studied by employing information-theoretic measures to detect global and local patterns and to measure the information content of graphs and networks. Fig.~\ref{figure1} shows the flowchart of the proposed testbed for assessing information loss/preservation in network dimensionality techniques. First, as a proof of concept, Fig.~\ref{figure2}A shows that the Shannon entropy of the adjacency matrix diminishes in value for a growing number of disconnected nodes. Fig.~\ref{figure2}B shows the impact of adding  disconnected nodes to a graph as an estimation error of approximations to graph entropy ($H(G)$), and of graph algorithmic complexity estimated by BDM ($BDM(G)$). The Block decomposition method (BDM) is a novel technique for approximating Kolmogorov complexity by means of algorithmic probability (c.f. Section~\ref{kmotifs}). Both BDM and $H$ measures behave as expected: while algorithmic complexity increases marginally due to the small information content added, with diminishing impact, by the contribution of every disconnected node, entropy asymptotically moves towards 0. Since the graph entropy and complexity are measured over the adjacency matrix of the graph, adding disconnected nodes means adding rows and columns of 0s, which are highly compressible and of low entropy and block entropy (entropy rate, i.e. taking as unit \textit{micro-states} all submatrices of $n\times n$ bits from $n=1$ to the length of the adjacency matrix). It follows then that algorithmic complexity captures important features of these graphs.

In~\cite{zenilgraph}, we showed that Deflate and BDM very closely approximated the complexity of dual graphs. Here we performed a similar test using cospectral graphs, with a surprising positive outcome. In graph theory, the set of eigenvalues of the adjacency matrix of a graph is referred to as the \textit{spectrum} of the graph. Two graphs are \textit{isospectral} or \textit{cospectral} if the adjacency matrices of the graphs have equal multisets of eigenvalues, i.e., the same spectra. Cospectral graphs may look very different; two examples are shown in~\ref{figure2}B. However, entropy (Fig.~\ref{figure2}B) and algorithmic complexity estimated by Bzip2 (Fig.~\ref{figure2}B and C) and BDM (Fig.~\ref{figure1}D) provided the same information content values for almost all co-spectral graphs considered. BDM (Fig.~\ref{figure2}D) provided better estimations (with higher rank correlation and less outliers) than Bzip2 and Entropy (Fig.~\ref{figure1}C) for 180 graphs and their cospectrals, that is, the original graphs and their cospectral counterparts had values closer to each other. This is consistent with the fact that cospectral graphs share important algebraic properties and should therefore have a similar information content, but it was not necessarily theoretically expected, there being no known procedure for producing all graphs with a certain spectrum and no simple algorithm for producing a cospectral graph from a given graph. In general, there is no one-to-one correspondence, and in this sense the cospectral information-content similarity is more surprising than that of dual graphs. That classic entropy, Bzip2 complexity and BDM based on algorithmic probability produce very similar complexity values for cospectral graphs means that these methods are (from worse to better) able to capture fundamental algebraic properties shared by cospectral graphs and so can be used, as we claim, for comparing reduction methods. 

As part of the dataset to be considered, we assessed the amount of information (in bits) in six networks from an Artificial Gene Network database: two networks with small-world (SW) topology, two scale-free networks (SF) and two Erd\"{o}s-R\'enyi (ER). In the past, most of the work on the complexity of graphs was focused on random networks, the so-called Erd\"{o}s-R\'enyi networks. But most of the interesting features of biological networks arise from the fact that these networks are not like random graphs. Connections among elements in a biological network are neither simple nor random. The small-world property of networks--signified by a small diameter--has been established beyond a doubt, revealing the key role of short cuts common in many real networks, from protein interactions to social networks, and from the network of hyperlinked documents to the interconnected hardware behind the Internet. Real networks, including biological networks, are also known to be scale-free~\cite{barabasi,barabasiscalefree2}. This suggests other possible mechanisms that could be guiding network formation.

Here we explore the complexity of these three large random graph classes, i.e., ER, SF and SW and various real-world, biological and non-biological networks. The results of the estimation of the Kolmogorov complexity of these artificial networks show that while there is no agreement as regards whether SW is more complex than SF or vice versa, for Shannon entropy, SW networks display greater combinatorial complexity (not shown in graphs). But for BDM, SF networks are more complex (Fig.~\ref{figure3}A,B), and both compression algorithms are in agreement as to the slightly greater complexity of SW and ER networks (Fig.~\ref{figure3}A,B). And in fact we have found that both BDM and compression can separate these graph in topological groups (see~\cite{zenilgraph} and~\cite{zenilkianitegner}). However, compression algorithms reverse the complexity order among SF, SW and ER, which is once again in agreement with BDM on motifs as a network dimensionality reduction method (Fig.~\ref{figure5}), thus showing that BDM does not harbor a bias toward motifs. That compression of the original graphs retrieves a different order than BDM and compression on motif profiles is counterintuitive because SW networks for small rewiring probability are very close to regular (ring/cycle) networks and should therefore not have large complexity values. However, compression algorithms differ from BDM in that they are entropy rate estimators and can therefore be fooled if no trivial statistical regularities are found.

Since we have normalized Kolmogorov complexity estimations by number of edges and nodes, this result can be compared directly with other networks, and we do not need to have exactly the same number of nodes or edges for comparison. Fig.~\ref{figure3} shows the complexity values and information content estimations of the 16 graphs from~\cite{alon} and the Mendes DB~\cite{mendes} using Bzip2 and Deflate lossless compression algorithms (BDM cannot be applied directly to real-number values, see Section~\ref{kmotifs}) as approximations to Kolmogorov complexity normalized by node.

Interestingly, we see BDM values retrieve differences between networks, meaning that local regularities better characterize them. So BDM values can be used to characterize families of networks.

\subsection{Information loss and conservation in network reduction methods}

We report the results of our evaluation of the loss and preservation of information in network reduction techniques. To do this we first measure the information content of the adjacency matrix of a graph $G$, then the information content of the graph $G^\prime$ resulting from the application of each dimensionality reduction method. Finally we consider the difference of these values $D_{C,M}(G, G^\prime) = C_{M}(G) - C_{M}(G^\prime)$ for complexity measure $C$ and reduction method $M$. In general, $D_{C,M}(G, G^\prime)<0$, but some methods, such as spectral analysis (c.f. Section~\ref{kspectra}), can lead to the introduction of spurious information such that $D_{C,M}(G, G^\prime)>0$, especially for complexity measures of an entropic nature, such as compression (in contrast to those of an algorithmic nature such as BDM). But we are mostly interested in the case in which given 2 graphs $G_1$ and $G_2$ such that $C(G_1) < C(G_2)$ for complexity measure $C$, then $C_{M}(G_1^\prime) < C_{M}(G_2^\prime)$, especially for cases in which this is preserved across different $C$s.

The subgraph complexity (BDM) and lossless compression (Bzip2) values of the networks (Fig.~\ref{figure3}) that were classified by their network motifs have been studied before~\cite{superfamilies}, in order to assess the preservation of relative information content. That is, whether $C_{M}(G)<C_{M}(G^\prime)$, where $M$ is any of the complexity methods used in this project: BDM, Deflate (Compress) and Bzip2, on all reduction methods: motif profiles, graph spectra and sparsification. The results summarized in Fig.~\ref{figure3} encompass genetic, protein, power grid and social networks, as described in~\cite{superfamilies}. The plot shows that compression and BDM preserve to some degree the relative information content of most types of networks but BDM produces a convex curve while all others are more concave (Fig.~\ref{figure3}B). While Deflate and Bzip2 show different degrees of success at distinguishing families of networks, BDM was the best at distinguishing networks by their families assigning lower or higher complexity to different groups (e.g., genetic vs protein vs electric, or Erd\"os-R\'enyi vs scale-free vs small world) even normalizing by edge density and thus truly capturing essential differences of their topological properties. This is consistent with the main result in~\cite{milo,superfamilies}, showing that local graph structures can classify network families with great precision, BDM, however, looks at local structures in the network adjacency matrices instead, which is a proper superset than counting subgraphs (motifs) as done before in the cited papers.

\subsubsection{Loss of information quantification in graph spectra}

\label{kspectra}

There is an extensive literature on connecting graph properties to the eigenvalues of graph adjacency matrices. The so-called eigenvalue spectrum of these graphs provides information about their structural properties. Eigenvalues store information about a graph. Many properties of a graph can be recognized from its spectrum~\cite{newman}. 
We have calculated the amount of information preserved in spectra of different network families. Graph spectra can characterize certain properties of graphs. For example, spider graphs with $k$ rays have redundant eigenvalues, and the spider graph spectrum characterizes the graph by its number of rays and diameter. Indeed, it follows from the configuration of the adjacency matrix that the spectrum of a spider graph of $k$ rays and diameter 1 is: 

$(-\sqrt{k}, \sqrt{k}, \underbrace{0,\ldots, 0}_{k-1 times})$, 

with spiders of greater diameter having slightly greater complexity. This simplicity in the redundancy of the spectrum of spider graphs is consistent with their low Kolmogorov complexity.

Unlike the process of growing a spider graph, growing a random graph with edge density 0.5 requires a larger amount of information to specify the graph spectrum. Indeed, the Kolmogorov complexity of the spectrum of a spider graph is bounded by the number of rays $k$ with the same eigenvalue $e$ with $K$ complexity $2 \log(k) K(e)$, and the number $m$ of trailing 0s with Kolmogorov complexity $O\log(m-1)$. All biological networks were subject to the greatest loss of information when spectral sparsification was used (see Fig.~\ref{figure3}, where spectral curves are mostly flat, thus not allowing us to distinguish between different networks). This is because spectra analysis is lossy (many graphs can have the same graph spectrum) and therefore is bound to lose vital information, even if spectra capture important algebraic properties of a network. Biological networks were also found to have close to nilpotent eigenvalues, but we found no theoretical explanation for this (see Fig.~\ref{figure4} where biological networks have values closer to $x=0$). We think the reason is the high number of low degree nodes in biological networks. Indeed, it has been pointed out that the spectrum of these networks is quite susceptible to fluctuations of the vertex degrees, and in the case of irregular graphs the eigenvalues of the adjacency matrix just mirror the tails of the degree distribution and thus do not reflect any global graph properties~\cite{gundert}.

While spectra analysis is known to be a lossy data reduction technique, our results show that spectra analysis respects the information order of real-world networks, as compared to the full lossless compressed lengths of the networks. Another interesting phenomenon was the perfect match of values between BDM and Deflate for the synthetic networks. Thus, taken together, BDM and Deflate perfectly differentiate between the natural and artificial networks to be further investigated. That graph spectra are inconsistent with the common estimation trend of Kolmogorov complexity, as reported in previous experiments, suggests that graph spectra analysis is the method with the greater loss of information. Yet this does not make it less interesting as a measure for quantifying certain aspects of a graph, provided we take into account that this method may indeed lose the relative complexity and information content of the original graph.

The graph Laplacian may be claimed to more naturally represent some properties of graphs, when compared to the plain graph spectrum. From the point of view of information content, the Laplacian cannot contain more information than the information that can already be extracted from the adjacency matrix. Indeed, the Laplacian is defined as $L = D - A$, where $D$ is a diagonal matrix where each diagonal entry is the number of links for each node and $A$ the adjacency matrix. $D$ can clearly be derived from $A$ as the sum of 1s in each row. Moreover, the calculation of the Laplacian is of fixed size. Hence $K(A)=K(L)+c$ differs only by a constant value. But it remains to be ascertained whether the Laplacian conserves more information than the regular graph spectrum, despite the fact that both retrieve the same number of vector entries.

Taking the information content from the spectra alone does not preserve the relative order or show any clustering capabilities by type of network. This means that when using BDM, graph motif and compressibility analysis, order is better preserved among networks of the same family than among different families.

\subsubsection{Network motifs preserve local information and characterize graphs}

\label{motif}

The idea of a local scale subgraph-based analysis was first presented in~\cite{milo}, when network motifs were discovered in the gene regulation (transcription) network of the bacteria E. Coli and then in a large set of natural networks. A network motif is defined as a recurrent and statistically significant sub-graph occurring in a network or across various networks. More formally, if $G$ and $H$ are two graphs, $H \subset G$, the number of appearances of graph $H$ in $G$ is referred to as the frequency $F_H$ of $H$ in $G$. A graph is referred to as recurrent (or frequent) in $G$ when its frequency $F_G(G^\prime)$ is above a predefined threshold or cut-off value (usually compared to a random graph).

Much work has been done on the subject, resulting in the discovery of characteristic motifs among species and network types, and even superfamilies~\cite{superfamilies} of network motifs that characterize complete classes of networks such as transcription interaction, signal transduction, even social networks. Motifs have recently garnered much attention as a useful concept for uncovering the structural design principles of complex networks~\cite{superfamilies}. There have been suggestions that motif analysis cannot deliver on the promise of a deeper understanding of the networks studied (eg~\cite{GRN}), mainly because of a loss of information pertaining to context, i.e., the broken connections between subgraphs~\cite{knabe}. While it is clear that local scale information is lost, it is not clear how much a subgraph analysis can preserve of the information content of the original full-size networks. Motifs have been of signal importance largely because they may reflect functional properties. We ask how much information can be recovered by looking at a network on  a very local scale, as proposed by the network motif analysis approach. The concept of algorithmic probability will enable us to approximate and add small-scale complexity from the decomposition of a network into its possible subgraphs in order to determine the amount of information that is preserved in this bottom-up approach, as compared to the information content of the full-size network.

In Fig.~\ref{figure2} we show the motifs, as calculated by the open-source software FANMOD~\cite{fanmod}, of Escherichia coli~\cite{ecoli}, together with information-theoretic measures associated with each motif. We see that Shannon's entropy distinguishes two cases, assigning the two lowest possible entropic measures ($\log(2)$ and $\log(3)$), while BDM approximations provide a finer-grained classification, retrieving 3 different values for all 4 motifs. Both Shannon entropy and Kolmogorov complexity approximations agree on the equal complexity of the first two motifs. Results of applying compressibility (Deflate and Bzip2) and algorithmic probability (BDM) to approximate the Kolmogorov complexity motifs of the artificial network showing the agreement of the compressed size of network motif files, network motifs of size 4 and 5, when compared to the complexity of the original networks (BDM, Bzip2 and Deflate) (see Fig.~\ref{figure3}).

Fig.~\ref{figure3}C shows that natural and synthetic networks that belong to the same family or have the same topology have similar complexity values, both for the original and for the motif compressed file sizes. The same compression trend is confirmed between motifs and BDM for both sets of graphs, providing further evidence of the connection established in this paper between the information content of subgraphs (more properly, some subarrays of the adjacency matrices) and the frequency of a subset of overrepresented graphs (known as graph motifs). Similarly, but to a lesser degree, Bzip2 and Deflate (Fig.~\ref{figure3}A-B) show network family clustering capabilities, assigning graphs of similar origins or topology more or less the same incompressibility values as approximations of their complexity/information content.

\subsubsection{Preserved and lost information in network sparsification}

\label{sparse}

Sparsification can be viewed as a procedure for finding a set of representative edges and weighting them appropriately in order to choose a smaller but representative number of vertices and edges that preserve important features of a network, for example, its \textit{backbone}. Sparse graphs are easier to handle than denser ones and can be used for network dimensionality reduction for the study of very large networks. A sparse graph is one whose number of edges is reasonably viewed as being proportional to its number of vertices. One may consider a graph sparse if its average degree is less than 10~\cite{batson}. While real-world networks are already sparse by most standards, because of their typically large size it is often useful to reduce their dimensionality further in order to enable inspection of the most important connections, for example, in biology, where even a new link of regulation between genes can be a breakthrough. Sparsification methods have been used in biology (eg~\cite{tsuda,august}). It has traditionally been shown that these algorithms preserve topological properties of the original networks after sparsification, but little is known about the information content conservation. In this section we calculated the amount of information preserved in spectral sparsifiers of different types of network. We used the algorithm suggested in~\cite{spielman} for the purpose, a fast algorithm to calculate sparse networks by random sampling, where the sampling probabilities are given by the effective resistances of the edges. The effective resistance of an edge is known to be equal to the probability that the edge appears in a random spanning tree of $G$. It has been proved that for each error parameter $\epsilon$ there is such a spectral sparsifier, and that it can be calculated in $n\log n/\epsilon$ time for some large constant $n$--independent of the sampling method--by replacement from graph $G$.

$Spar(G)$ will denote the graph resulting from the application of the sparsification method to $G$. Here we are interested in determining whether this other method actually preserves the information of the network, beyond topological properties. To which end we again measure the information content--by way of Shannon Entropy and Kolmogorov complexity--of the networks previously studied. Figs.~\ref{figure3} show that the method does indeed follow the relative information content of the lossless compressed lengths of the original networks. We have chosen the error terms for all networks so as to keep 20\%, 40\%, 60\% and 80\% of the edges, following a recent, widely accepted network sparsification algorithm, as described in~\cite{spielman}. We report the findings for the rate of information loss in Fig.~\ref{figure3}. The information loss rates for sparsification preserving degree distribution (see Fig.~\ref{figure5}) (differences between 20\%, 40\%, 60\% and 80\% threshold values) are -2.44, -0.908 and -0.611. The relative order of information content was preserved upon application of all methods. Only Bzip2 reports an inconsistency in the relative information conservation for SW networks. The rest--including Deflate--indicate good preservation of the features that characterize the information content of the original networks.

We calculated the graph spectra of several real-world networks from~\cite{milo}. $Spec(G)$, consisting of a list of eigenvalues of the adjacency matrix of $G$, denotes the spectra. Fig.~\ref{figure3} shows the result of compressing both the original networks and their graph spectra. The approximation of the algorithmic information content preserved by the spectra is calculated by losslessly compressing $Spec(G)$ of $i$ eigenvalues of the graph adjacency matrix of $G$ of size $i$, and is denoted by $C(Spec(G))$, where $C$ is a lossless compression algorithm (e.g., Bzip2 or Deflate) and $Spec(G)$ the eigenvalues sorted from smallest to largest. As seen in Fig.~\ref{figure3}, BDM fully characterizes network topology (see the synthetic network values) and assigns similar complexity to similar networks, in agreement with previously reported motif analysis results.

Fig.~\ref{figure3} shows the complexity values for the protein networks 1, 2 and 3; social networks 1 and 2; electronic circuit networks 1 and 2; genetic networks (yeast and ecoli); and 3 types of graph with different topologies (Erd\"os-R\'enyi, scale-free and small-world from the Mendes DB). The rate of information loss is clear, with the greatest loss at 80\% and then diminishing at a decreasing speed the greater the sparsity, keeping relative information but deleting edges at the determined values. Trends show that the algorithm preserves the absolute and relative information content of the original networks.

Fig.~\ref{figure4} shows a very interesting phenomenon. Reaching a 40\% sparsification value has the diametrically opposite effect to losing information; the resulting network appears more random because most of the structure is lost. Then at 20\% the original trend resurfaces; the resulting sparse graph is truly small as compared to $> 60\%$ and comes last, with the smallest information content. Combined, this strongly suggests that keeping less than 50\% leads to important information being lost, and some complexity may actually be introduced (e.g., from graph disconnection). This of course depends on the topological structure of the graph--it is known that scale-free networks are more robust in the face of random failure but less so in the face of targeted attacks~\cite{albert}. This is in contrast to motif analysis, as shown in~ Subsection~\ref{kmotifs}, where it was demonstrated that very few elements of local structure (subgraphs) preserve the basic information necessary to continue characterizing the networks. Sparsification is thus seen to be safe for real-world networks at a 50\% value, and unsafe for lower values, where most of the information begins to be lost, as happens in spectral analysis.

\section{Conclusions}

While a variety of dimensionality reduction techniques have been proposed in recent years, beyond network motif analysis and sparsification techniques, there has not been much done in the area of network dimensionality reduction. Here, we presented a novel and systematic way to compare old and new dimensionality reduction techniques based on information theory. The suggested methodology is based on the calculation of the preservation of information content. Here it was empirically demonstrated the application of these novel methods. We have measured their effectiveness on a relatively limited but representative set of data, and reported their potential and associated information loss for dimensionality reduction. While our empirical results are a useful pointer, further numerical and theoretical work is probably needed to understand better the reasons underpinning the reported results. 

As a proof of concept, we first showed that approximations of the information content of cospectral networks are similar, as is consistent with the theory. We then tested three important graph dimensionality reduction techniques, showing the various ways and the degrees to which each method is capable of preserving the information content of the original networks. We calculated for the first time the impact of applying three important network reduction techniques to the information content of the 3 most important random and complex graph models, namely Erd{\"o}s-R\'enyi, Barab\'asi-Albert and Watts-Strogatz.

The results of the experiments reveal that the sparsification method evaluated preserves relative information and that its rate of information loss is as expected, but in deleting more than half the edges it leads to significant inconsistencies and loss of information. In the case of motif analysis, we found results in agreement with the method based on algorithmic probability that approximated the algorithmic information content of a network by considering local regularities, validating (and generalizing) the surprising fact that local regularities (subgraphs) preserve information to such a degree that important profiling information from the networks is fully recoverable (e.g., their type across different superfamilies), as reported in~\cite{milo}. Finally, graph spectra was the most irregular reduction technique, capturing important algebraic information in a condensed fashion but in the process losing most of the information content of the original network.

The results we report indicate that a local complexity approach retrieves enough local information about networks to distinguish between families, which is not possible by averaging their information content on the global scale through applying lossless compressibility to the complete networks. The results suggest that despite its local nature, motif analysis is the method that preserves the most information,  while sparsification techniques are to be used carefully and cannot reduce the network edge density by more than 50\% without losing information essential for characterizing the network's original complexity. And finally, while graph spectra analysis captures important algebraic features, it is to be used with the greatest care, as it is definitely the technique that loses most of the original information content, making it impossible to reconstruct properties of the original graph in the general case. The paper explains these results by identifying weaknesses among these techniques and providing instructions on what they are best at and what to avoid, thus making it possible to improve the application of these methods for different purposes and clearing a path to assess other techniques and make meaningful comparisons. It helps to evaluate and compare network dimension reduction techniques that have been proposed so far and may be introduced in the future.




\section{Acknowledgments}

The authors would like to thank our Karolinska Institutet colleagues: Gordon Ball for his valuable technical help, and the other members of the Unit of Computational Medicine. This work was supported in part by the Foundational Questions Institute (HZ), the John Templeton Foundation (HZ), the VINNOVA (VINNMER) Marie-Curie fellowship, (NK), AFA Insurance (JT), Torsten S{\"o}derberg Foundation (JT), STATegra (JT, HZ), the Stockholm County Council and the Swedish Research Council. The funders played no role in the design of the study, in data collection and analysis, in the decision to publish, or in the preparation of the manuscript.

\section*{Appendix}

\subsection{Network dimensionality reduction techniques}

\subsubsection{Network spectral sparsification}

Network dimensionality methods have been introduced and used in biology ~\cite{tsuda,august,milo,zenilgraph} for purposes such as analysis and profiling. In general the goal of network sparsification is to approximate a given graph $G$ by a sparse graph $H$ on the same set of vertices. If $H$ is close to $G$ in some appropriate metric, then $H$ can be used as a signature, preserving important properties of $G$ for faster computation after reducing the size of $G$ and without introducing too much error, thus making computation time and storage of $H$ cheaper, as the network is more sparse compared with $G$. Obvious trivial sparsification methods include edge deletion by some criteria, such as the outermost ones (called a $k$-shell~\cite{shell1,shell2}, often used to identify the core and the periphery of the networks), but most of them (such as the aforementioned) are rather arbitrary or ad-hoc, devised for specific purposes, rather than general methods aiming at preserving important, algebraic, topological or dynamical properties of the original graph.

Several notions of graph sparsification have been proposed. For example, a method motivated by proximity problems in computational geometry was introduced in the form of \textit{graph spanners}~\cite{chew}. A spanner is a sparse graph in which the shortest-path distance between every pair of vertices is approximately the same in the original graph as in the spanner. For example, a popular sparsification algorithm is the \textit{spanning tree}~\cite{graham} designed to preserve node distance but clearly destroy all other local node properties such as the clustering coefficient.

Not many non-trivial methods for network dimensionality reduction exist today, and it is acknowledged~\cite{spielman,spielman2,batson} that spectral graph sparsification is among the most efficient both in preserving important algebraic and dynamical properties of a network and in terms of fast calculation. In part the lack of more methods is due to a lack of assessment tools using which to decide whether one method is better than another in general terms, rather than whether it preserves one or another specific graph theoretic property (e.g., the transitive edge deletion method destroys the clustering coefficient of the original graph~\cite{aho}). Among the methods considered in this paper is a high-quality cutting-edge one~\cite{batson,spielman} based on graph spectra.

Graph spectral sparsification is a technique that has been used in data analysis as a dimensionality reduction method in biology~\cite{banerjee}. However, whether most graphs are uniquely determined by their spectrum is an open problem. But because at least some graphs share the same spectrum the process is lossy, because one cannot fully recover the original graph from its spectrum, at least in these cases. For example, almost all trees are cospectral (the share of cospectral trees on $n$ vertices tends to 1 as $n$ grows), where \emph{almost} means that the set of elements for which the property does not hold is a set of measure zero.

A good introduction to spectral graph sparsification may be found in~\cite{batson} and we use it to illustrate the network dimension reduction assessing tool introduced in this paper. The notion upon which all these methods are based is the spectral similarity of graph Laplacians. Spectral sparsification requires that the Laplacian quadratic form of the sparsifier approximate that of the original graph on all real vector inputs. This is equivalent to saying that the Laplacian of the sparsifier is a good preconditioner for the Laplacian of the original~\cite{liu}.

\subsubsection{Network motif profile analysis}

Another more recent method closer to biology works by looking at the subgraphs (of very small size) that make up a graph $G$. The method, introduced in~\cite{milo}, turns out to be capable of characterizing networks by the families to which they belong  (e.g., genetic versus social) and is therefore also a perfect candidate for quantifying the amount and type of information that is preserved and lost when retaining only the network motifs of the original graph. It compares the frequency of small subgraphs with randomized versions of the same network (i.e., networks with the same size and the same degree distribution). Over and under-represented subgraphs are called network motifs and turn out to characterize a network type.

Each of these network motifs, defined by a particular pattern of interaction between vertices, may reflect a framework in which particular functions are achieved efficiently. It is generally believed that motifs are of signal importance largely because they may reflect functional properties. The calculation of network motifs may provide a deep insight into a network's function but their calculation is computationally challenging. We are therefore limited to small sizes and hence to considering only local structures. The surprising result is that these local structures contain enough information about a system to characterize it uniquely, at least in the case of graphs with similar topologies and functions.

A graph in $G$ is considered frequent and therefore denoted as a motif when its frequency $F_G(G^\prime)$ is above a predefined threshold or cut-off value. There is an ensemble $\Omega(G)$ of random graphs corresponding to the null-model associated with $G$. We should choose $N$ random graphs uniformly from $\Omega(G)$ and calculate the frequency for a particular frequent sub-graph $G^\prime$ in $G$. If the frequency of $G^\prime$ in $G$ is higher than its arithmetic mean frequency in $N$ random graphs $R_i$, where $1 \leq i \leq N$, we consider this frequent subgraph `significant' and hence treat $G^\prime$ as a network motif of $G$. The $Z$-Score has been defined by the formula,

$$Z(G^\prime)=\frac{F_G(G^\prime)-\mu_R(G^\prime)}{\sigma_R(G^\prime)}$$

\noindent where $\mu_R(G^\prime)$ and $\omega_R(G^\prime)$ stand for mean and standard deviation frequency in set $R(G)$. The larger the $Z(G^\prime)$, the more significant is the sub-graph $G^\prime$ as a motif~\cite{milo}. The biological studies endeavor to interpret the motifs detected for biological networks. For example, the network motifs found in E. coli were discovered in the transcription networks of other bacteria such as yeast, among others~\cite{yeast}.

\subsection{Estimation of information content of a graph}

To approximate Kolmogorov complexity we use a technique based on algorithmic probability (see Section~

\ref{kmotifs}) and two universally employed lossless compression algorithms, Bzip2 and Deflate, the former set to maximum compression (option flag set at -9) and the latter in the default position as implemented in  \textit{Mathematica}'s Compress function version 10. Bzip2 is an open source data compressor that uses a stack of different algorithms superimposed one atop the other, starting with run-length encoding, Burrows-Wheeler or the Huffman coding, among others. We sometimes compare, strengthen or complement findings by also providing estimations of Shannon's information entropy on the adjacency matrix.

\subsubsection{Shannon Entropy of a graph}

\label{kmotifs}

Central to information theory is the concept of Shannon's information entropy~\cite{shannon}, which quantifies the average number of bits needed to store or communicate a message. Shannon's entropy determines that one cannot store (and therefore communicate) a symbol with $n$ different symbols in less than $\log(n)$ bits. In this sense, Shannon's entropy determines a lower limit below which no message can be further compressed, not even in principle. For an ensemble $X(R, p(x_i))$, where $R$ is the set of possible outcomes (the random variable), $n = |R|$ and $p(x_i)$ is the probability of an outcome in $R$. The Shannon information content or Entropy of $X$ is then given by

$$H(X)= - \sum_{i=1}^n p(x_i) \log_2 p(x_i).$$

To calculate the Shannon entropy of a graph $H(G)$, let the message $X$ be the adjacency matrix of a graph $G$ denoted by $Adj(G)$, then

$$H(Adj(G))= -\sum_i^{|G|\times|G|} p(Adj(G)) \log_2 p(Adj(G)).$$

\noindent where $|G|$ is the vertex count of the graph $G$ and the probability distribution is over the bits of the adjacency matrix.

For example, a complete graph and a completely disconnected graph would have minimal Shannon entropy because the adjacency matrix entries are either all 0 or all 1 (assuming self-loops). However, Erd\"os-R\'enyi (ER) graphs with edge density 0.5 would have maximal Shannon entropy because their adjacency matrices have about the same number of 1s and 0s and are therefore statistically `typical'-- every bit is equally highly surprising. In other words, the bits of the adjacency matrices of complete and completely disconnected graphs are unsurprising, because getting a 1 after a long list of 1s, or a 0 after a long list of 0s does not add any Shannon information.

This is, however, very different in algorithmic information theory, where we are interested in whether bits are causally related. For example, the adjacency matrix of a directed complete graph--the direction being that the matrix is diagonal, with either all 1s on one side of the diagonal and 0s on the other side (or the other way around)--would have maximal Shannon entropy but is clearly not random, and should therefore have a low (algorithmic) information content. We therefore used a graph algorithmic complexity measure rather than this statistical combinatorial one.

\subsubsection{Graph Algorithmic Probability}

\label{kmotifs}

The concept of algorithmic probability (also known as Levin's semi-measure) yields a method for approximating Kolmogorov complexity related to the frequency of patterns in the adjacency matrix of a network, including therefore the number of subgraphs in a network. The algorithmic probability~\cite{solomonoff,levin,chaitin} of a subgraph $H \in G$ is a measure that describes the probability that a random computer program $p$ will produce $H$ when run on a 2-dimensional tape universal (prefix-free\footnote{The group of valid programs forms a prefix-free set (no element is a prefix of any other, a property necessary to keep $0 < m(G) < 1$). For details see~\cite{calude}.}) Turing machine $U$. That is, $m(G) = \sum_{p:U(p) = H \in G} 1/2^{|p|}$. An example of a popular 2-dimensional tape Turing machine is Langton's ant~\cite{langton}, commonly referred to as a \textit{Turmite}.

The probability semi-measure $m(G)$ is related to Kolmogorov complexity $K(G)$ in that $m(G)$ is at least the maximum term in the summation of programs ($m(G)\geq2^{-K(G)}$), given that the shortest program carries the greatest weight in the sum. The algorithmic Coding Theorem~\cite{cover} further establishes the connection between $m(G)$ and $K(G)$ as (\cite{levin}): $|-\log_2 m(G) - K(G)| < c$ (Eq. 1), where $c$ is some fixed constant, independent of $s$. The theorem implies that~\cite{cover,calude} one can estimate the Kolmogorov complexity of a graph from its frequency of production by running random programs that simply rewrite Eq.~(1) as: $K(G^\prime)=-\log_2 m(G) + O(1)$.

In~\cite{d4} a technique was advanced for approximating $m(G)$ (hence $K$) by means of a function that considers all Turing machines of increasing size (by number of states). Indeed, for small values of $n$ states and $k$ colors (usually 2 colors only), $\textit{D}(n, k)$ is computable for values of the Busy Beaver problem~\cite{rado} that are known, providing a means to numerically approximate the Kolmogorov complexity of small graphs, such as network motifs. The Coding theorem then establishes that graphs produced with lower frequency by random computer programs have higher Kolmogorov complexity, and vice versa. The method is called the \textit{Block decomposition method} (BDM) because it consists of decomposing the adjacency matrix of a graph into subgraphs of sizes for which complexity values have been estimated, then reconstructing an approximation of the Kolmogorov complexity of the graph by adding the complexity of the individual pieces according to rules of information theory, as follows:

\begin{equation}
\label{newecaeq}
K(G) = \sum_{(r_u,n_u)\in Adj(G)_{d\times d}} \log_2(n_u)+K(r_u)
\end{equation}

where $Adj(G)_{d\times d}$ represents the set with elements $(r_u,n_u)$, obtained when decomposing the adjacency matrix of $G$ into all subgraphs contained in $G$ of size $d$. In each $(r_u,n_u)$ pair, $r_u$ is one such submatrix of the adjacency matrix and $n_u$ its multiplicity (number of occurrences). As is evident from the formula, repeated subgraphs only contribute to the complexity value with the subgraph BDM complexity value once plus a logarithmic term as a function of the number of occurrences. This is because the information content of subgraphs is only sub-additive, as one would expect from the growth of their description lengths (``$n$ times a subgraph''). Applications of $m(G)$ and $K(G)$ have been explored in~\cite{d4,d5,numerical,kolmo2d}, and include applications to graph theory and complex networks~\cite{zenilgraph} and~\cite{kolmo2d} where the technique was first introduced.

\subsubsection{Calculation of graph motifs}

In Fig.~\ref{figure2} the motif-analysis software called FANMOD~\cite{fanmod} was used to calculate the graph motifs (the over-represented subgraphs), and we took the output files with the adjacency matrices in string form. This was done for motifs of size 4 and 5. The files considered contained the occurring subgraphs in string notation followed by their frequency of occurrence, so in a strict sense these files are already a compressed version as they only contain the different subgraphs but not their repetitions, other than as encoded in their frequencies. In the files only motifs were considered, that is, subgraphs of size 4 and 5 that were either over or under-represented as compared with randomized versions of the same network. More precisely, motifs were calculated with FANMOD by using a parameter absolute $Z$ score larger than 2, a $p$-value less than 0.05, a frequency of at least 0.01\%, and included in the output file motifs that have been found at least 5 times.

The files were therefore further compressed with both Bzip2 and Deflate in order to capture possible statistical regularities in the type and frequencies of the motifs. Then the files were compared to the compressed lengths of the original networks for both compression algorithms. The underlying rationale is that non-random graphs will show an over-representation of certain motifs and an under-representation of others, hence reducing or increasing the number of these objects in the resulting files. Indeed, from the output files of FANMOD one can reconstruct some of the information of the original network, with the number of subgraphs and their frequency, but some information is lost in the form of the way in which all these subgraphs (motifs) may have been connected. Motif analysis displays both conservation of information and clustering capabilities by families, as reported in~\cite{superfamilies} and verified again here with BDM. Results are summarized in Fig.~\ref{figure2}C. It is worth noting that because BDM looks at local regularities only, it may be biasing or amplifying the results toward network motifs over other network dimensionality reduction approaches. This is not a problem but nonetheless something to be taken into consideration. Another interesting phenomenon to investigate is the information loss and preservation when considering all possible induced subgraphs (graphlets) in a graph~\cite{derek}.

\end{document}